\documentclass[12pt,a4paper,oneside]{article}
\pdfoutput=1
\usepackage{graphicx,amssymb,amsmath,slashed}
\usepackage[margin=2cm]{geometry}

\newcommand{\eg}{{\it e.g.}}
\newcommand{\beq}{\begin{equation}}
\newcommand{\eeq}{\end{equation}}
\newcommand{\co}{\mathcal{O}}
\newcommand{\cL}{\mathcal{L}}

\newcommand{\afb}{A_{\rm FB}^{t \bar t}}
\newcommand{\mtt}{M_{t \bar t}}

\begin{document}


\title{\bf \Large Predictions from Heavy New Physics Interpretation of the Top Forward-Backward Asymmetry}

\author{\fontsize{12}{16}\selectfont C\'edric Delaunay$^1$, Oram Gedalia$^2$, Yonit Hochberg$^2$ and Yotam Soreq$^2$ \vspace{6pt}\\
\fontsize{11}{16}\selectfont\textit{$^1$ CERN Physics Department, Theory Division,
CH-1211 Geneva 23, Switzerland}\\
\fontsize{11}{16}\selectfont\textit{$^2$ Department of Particle
Physics and Astrophysics, Weizmann Institute of Science,}\\
\fontsize{11}{16}\selectfont\textit{Rehovot 76100, Israel}}
\date{}
\maketitle

\begin{abstract}
We derive generic predictions at hadron colliders from the
large forward-backward asymmetry observed at the Tevatron,
assuming the latter arises from heavy new physics beyond the
Standard Model. We use an effective field theory approach to
characterize the associated unknown dynamics. By fitting the
Tevatron $t\bar t$ data we derive constraints on the form of
the new physics. Furthermore, we show that heavy new physics
explaining the Tevatron data generically enhances at high
invariant masses both the top pair production cross section and
the charge asymmetry at the LHC. This enhancement can be within
the sensitivity of the 8~TeV run, such that the 2012 LHC data
should be able to exclude a large class of models of heavy new
physics or provide hints for its presence. The same new physics
implies a contribution to the forward-backward asymmetry in
bottom pair production at low invariant masses of order a
permil at most.
\end{abstract}

\section{Introduction}

Physics beyond the Standard Model (SM) has thus far eluded
experimental discovery. The 7~TeV run of the LHC has not
produced any signs of new TeV scale states. Nevertheless, New
Physics (NP) around the TeV scale is required by naturalness of
the electroweak scale, which in the SM is mainly destabilized
by the heaviness of the top quark. It is thus sensible to
expect that a (technically) natural UV completion of the SM
extends the top sector, such that top quark observables at
hadron colliders might deviate from their SM predictions. In
this theoretical context, the observation of an anomalously
large forward-backward asymmetry in top pair production at the
Tevatron is intriguing.

At the inclusive level, both CDF and D0 have been consistently
reporting values exceeding the SM prediction by $\sim2\sigma$
in various channels~\cite{Aaltonen:2011kc,cdfdilAFB,
Abazov:2011rq,cdfAFBnew}. Most striking is the evidence for a
growth of the asymmetry with $\mtt$ reported by
CDF~\cite{cdfAFBnew}, which is well illustrated by their
measurement of the asymmetry above 450~GeV:
\beq \label{AFBhigh}
\afb(\mtt \geq 450\, \mathrm{GeV})=+0.296 \pm 0.059 \, ({\rm
stat.})\, \pm 0.031 \, ({\rm syst.}) \,.
\eeq
Although the central value is lower than their previous
result~\cite{Aaltonen:2011kc}, it is still $\sim3\sigma$ above
the SM prediction of $\sim0.1$~\cite{cdfAFBnew}. The latter
might be underestimated due to higher-order QCD and electroweak
effects~\cite{Kuhn:2011ri} (see also the impact of the
renormalization scale uncertainty~\cite{Brodsky:2012ik}), yet
the above series of observations is certainly worth exploring
in the context of NP beyond the SM.

There are two LHC measurements of direct relevance to the top
$\afb$: the $t \bar t$ differential cross section as a function
of $\mtt$ and the charge asymmetry ($A_{\rm C}$) in $t\bar{t}$
production. The 7~TeV LHC run has already provided preliminary
measurements of these two observables. For instance, the CMS
collaboration recently measured the charge asymmetry at the
inclusive level,
\beq \label{ACdata}
\left(A_{\rm C}\right)^{\rm inc}=+0.004 \pm 0.010 \, ({\rm
stat.}) \, \pm 0.011 \, ({\rm syst.}) \,,
\eeq
as well as its distribution with $\mtt\,$, based on
5.0~fb$^{-1}$ of data~\cite{cmsAC}. Similar results were
obtained by the ATLAS collaboration~\cite{ATLAS:2012an,atlasAC}. These
results are consistent with the SM expectations. Measurements
of the differential cross section at the LHC have been carried
out using conventional top-reconstruction techniques (see
\eg~\cite{cmsdif}), and good agreement with the QCD prediction
below $\mtt \lesssim 700$~GeV is found.

We interpret the discrepancy between the Tevatron $\afb$ data
and the SM expectations as originating from heavy NP. (For
previous relevant works see~\cite{Jung:2009pi,Degrande:2010kt,
Jung:2010yn,Blum:2011up,Delaunay:2011gv,AguilarSaavedra:2011vw},
and for a review of various NP models see
\eg~\cite{Kamenik:2011wt}.) We assume that the scale of the new
physics is well above the scale $\mtt$ that is relevant to the
Tevatron measurements, and is such that the heavy NP is not
produced on-shell at the LHC running at center-of-mass energies
up to 8~TeV. These effects are well captured by an Effective
Field Theory (EFT) expansion below the pole of the NP states,
and can therefore be studied completely generically without
specifying any UV completion. Moreover, the wait for signals of
NP at the LHC together with electroweak precision data from
LEP~\cite{Barbieri:2000gf, Barbieri:2004qk, ALEPH:2005ab} imply
that heavy NP at the few TeV scale is a motivated framework.

The main goal of the present paper is to demonstrate within the
EFT framework that the heavy NP explanation of the anomalous
$\afb$ is still largely consistent with the 2011 LHC
measurements, and that the LHC run at 8~TeV should have enough
sensitivity to observe the effects of the heavy NP or to
exclude most of the parameter space accommodating the Tevatron
data.

The paper is organized as follows. Section~\ref{sec:ops} lists
the relevant EFT operators. In Section~\ref{sec:fit} we perform
a global fit of the operators to the most updated Tevatron
$t\bar{t}$ data set. Section~\ref{sec:lhc} compares the result
of the global fit to present ATLAS and CMS data, and gives
predictions for the differential $t\bar{t}$ spectrum and charge
asymmetry at the LHC. In Sections~\ref{sec:di} and~\ref{sec:bb}
we comment on the possible interplay of the top data with dijet
production at the LHC and forward-backward $b\bar{b}$
production asymmetry at the Tevatron, respectively. We conclude
in Section~\ref{sec:conc}.

\section{Effective Field Theory Approach} \label{sec:ops}

In order to collectively describe heavy NP, we use a set of
effective operators, under the assumptions that the scale
$\Lambda$ characterizing NP is significantly higher than the
top pair invariant mass in the relevant measurements. These
operators must relate an initial $u\bar u$ state to a final
$t\bar t$ state, and as such appear at dimension six and
higher. Operators leading from $d \bar d$ to $t \bar t$ are
suppressed at the Tevatron due to the small $d \bar d$
luminosity. Since our goal is to use the Tevatron data to make
predictions for the LHC, these operators are henceforth omitted
from the analysis. For use of these operators to ameliorate
potential tension between Tevatron and LHC data
see~\cite{Drobnak:2012cz}.

The relevant dimension six operators are made of four-fermion
contact terms. The operators with vector and axial-vector
structure are:
\beq \label{AV_ops}
\begin{split}
\co^{1,8}_V &= \left(\bar{u}\gamma_\mu T_{1,8} u\right)
\left(\bar{t}\gamma^\mu T_{1,8} t\right)\,, \qquad \co^{1,8}_A
= \left(\bar{u}\gamma_\mu\gamma^5 T_{1,8} u\right)
\left(\bar{t}\gamma^\mu\gamma^5 T_{1,8} t\right)\,, \\
\co^{1,8}_{AV} &= \left(\bar{u}\gamma_\mu\gamma^5 T_{1,8}
u\right) \left(\bar{t}\gamma^\mu T_{1,8} t\right) \,, \qquad
\co ^{1,8}_{VA} = \left(\bar{u}\gamma_\mu T_{1,8} u\right)
\left(\bar{t}\gamma^\mu\gamma^5 T_{1,8} t\right)\,.
\end{split}
\eeq
The scalar and tensor operators are:
\beq\label{ST_ops}
\begin{split}
\co^{1,8}_S&= \left(\bar{u}\,T_{1,8}u\right)
\left(\bar{t}\,T_{1,8} t\right)\,, \qquad \co^{1,8}_P =
\left(\bar{u}\,T_{1,8}\gamma^5u\right)
\left(\bar{t}\,T_{1,8}\gamma^5t\right)\,, \\ \co^{1,8}_{SP}&=
i\left(\bar{u}\,T_{1,8}u\right) \left(\bar{t}\,T_{1,8}\gamma^5
t\right)\,, \qquad \co^{1,8}_{PS} =
i\left(\bar{u}\,T_{1,8}\gamma^5u\right)
\left(\bar{t}\,T_{1,8}t\right)\,, \\ \co^{1,8}_T &=
\left(\bar{u}\,T_{1,8}\sigma^{\mu\nu}u\right)
\left(\bar{t}\,T_{1,8}\sigma_{\mu\nu}t\right)\,,
\end{split}
\eeq
with $T_1\equiv 1$ and $T_8\equiv T^a\,$.  The operators
$\co_{V,A}^8$ in Eq.~\eqref{AV_ops} interfere with the SM and
thus contribute to $u \bar u \to t \bar t$ processes at
$\co(\alpha_s/\Lambda^2)$, while the rest of the operators in
Eqs.~\eqref{AV_ops} and~\eqref{ST_ops} only contribute at
$\co(1/\Lambda^4)$.

In principle, there are also chromomagnetic dipole operators
involving the gluon field strength which contribute to $u \bar
u \to t \bar t$. Their contribution at $\co(1/\Lambda^4)$ is
suppressed by at least $(m_t/\Lambda)$ compared to their
dominant $1/\Lambda^2$ effects~\cite{Delaunay:2011gv}. Their
interference level contribution has the same shape as the QCD
$\mtt$ distribution~\cite{Degrande:2010kt}, and so these
operators are strongly constrained by $t \bar t$ cross section
measurements at the Tevatron. We thus neglect the
chromomagnetic operators from further discussion. Dimension
eight operators interfering with the SM are subdominant
compared to the effects that we consider when the NP couplings
are sizable at the scale where NP is
on-shell~\cite{Delaunay:2011gv}, as is typically required by
the large $\afb$ measurement. As such, they are henceforth
omitted.

The total effective Lagrangian which we consider is then given
by
\beq \label{lagrangian}
\cL_{\rm eff}=\sum_i \frac{C_i}{\Lambda^2}\mathcal{O}_i \equiv
\sum_i c_i \mathcal{O}_i \,,
\eeq
where the sum is over the operators in Eq.~\eqref{AV_ops}
and~\eqref{ST_ops} and $c_i$ are of dimension [mass]$^{-2}$.
This effective description has a limited range of validity in
terms of energies. The range is limited from above by the
breakdown of the perturbative EFT description. Naive
dimensional analysis dictates $C_i \lesssim 16\pi^2$, and
plugging in the typical value for $c_A^8$ suggested by the
result of our fit (see Section~\ref{sec:fit}) leads to
\beq
\Lambda < \mathcal{O}(10\ \mathrm{TeV})\,.
\eeq
The lower edge of the validity range is around 1~TeV, above the
typical scale of the relevant observables.

In order to minimize the impact of next to leading order
corrections to these NP contributions, we normalize the latter
to the SM one in all calculations. We assume that the
$K$-factors are universal, so that the NP/SM ratios at leading
order and next to leading order are the same. For an extended
discussion, see~\cite{Blum:2011up,Delaunay:2011gv}.

\section{Global Fit to the Top Tevatron Data} \label{sec:fit}

The NP contributions described by the operators defined above
can be used to fit the Tevatron data on top related
observables. We begin by describing the relevant measurements.
We avoid the need to deal with correlations in the data by
considering only measurements which can be treated as
uncorrelated. We thus use the following Tevatron data and the
corresponding SM estimates:
\begin{itemize}
\item Inclusive $t \bar t$ cross section from
    D0~\cite{Abazov:2011cq} and the QCD NNLO
    calculation~\cite{Baernreuther:2012ws}.
\item Differential $t \bar t$ cross section as a function
    of $\mtt$ from CDF~\cite{Aaltonen:2009iz} (Table~3, all
    bins but the first), to be compared with the
    approximate NNLO calculation of~\cite{Ahrens:2010zv}
    (Fig.~12).
\item Inclusive $t \bar t$ forward-backward asymmetry as
    measured by D0 in the lepton+jet
    channel~\cite{Abazov:2011rq} and by CDF in the dilepton
    channel~\cite{cdfdilAFB}, and the corresponding NLO
    QCD+EW estimate quoted by CDF in~\cite{cdfAFBnew}.
\item Differential $t \bar t$ forward-backward asymmetry as
    a function of $\mtt$ from CDF and the NLO QCD+EW
    estimate~\cite{cdfAFBnew} (Table~XVII).
\end{itemize}

\begin{figure}[tb]
  \centering
  \includegraphics[width=.5\textwidth]{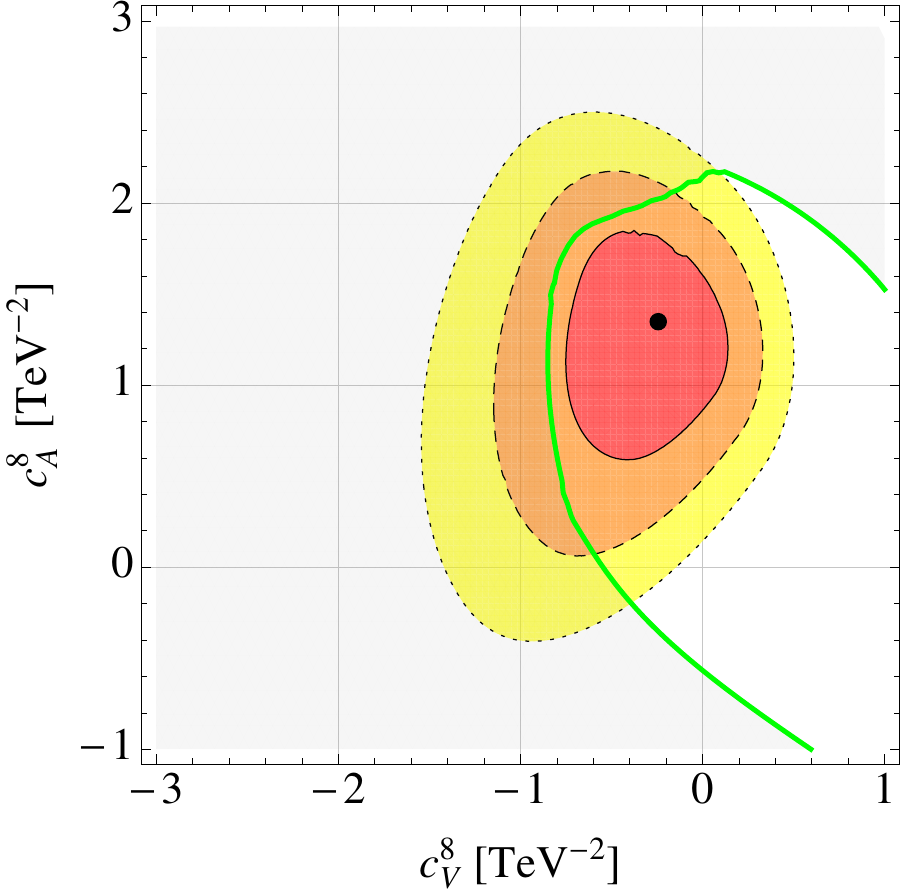}
  \caption{The result of the $\chi^2$ fit to the top related Tevatron data, presented in the plane
  of $c_V^8$ and $c_A^8\,$, while marginalizing over the rest of the operators. The red, orange
  and yellow shaded regions correspond to 1$\sigma$, 2$\sigma$ and 3$\sigma$, respectively,
  and the black dot stands for the best fit point. The light gray shaded region to the left of the
  green contour is excluded by the experimental bound on the $t \bar t$ cross section for $\mtt>1 \,
  \mathrm{TeV}$, Eq.~\eqref{Sbound}. \label{fig:chi2fit}}
\end{figure}

We fit the Wilson coefficients of the Lagrangian
Eq.~\eqref{lagrangian} to the Tevatron data listed above using
a $\chi^2$ analysis, where the observables are calculated using
the MSTW parton distribution functions~\cite{Martin:2009iq} at
leading order. Since large interference effects are required by
the data~\cite{Grinstein:2011yv} we derive confidence level
contours in the plane of the two interfering operators
$\co^8_{V,A}$ by marginalizing over all the non-interfering
operators. Note that in practice the full $t \bar t$
differential cross section contains only 6 independent terms,
and so the list of Wilson coefficients can be combined into 6
parameters. The result is presented in Fig.~\ref{fig:chi2fit}.
The best fit point corresponds to $c_A^8=1.35$~TeV$^{-2}$,
$c_V^8=-0.24$~TeV$^{-2}$ and mild contributions from the
non-interfering operators. The goodness-of-fit is 0.88.

We learn that heavy NP which does not interfere with the SM
($c_V^8=c_A^8=0)$ is in $\sim 3\sigma$ tension with the data.
Furthermore, the absence of the axial-octet operator $\co_A^8$
is in $\sim2\sigma$ tension with the data. Qualitatively, the
non-interfering operators do not relax the need for
interference mainly due to their rapid growth with $\mtt\,$,
which makes them strongly constrained by the measurement of the
differential cross-section in the hard regime.

If the heavy NP generates a single operator of definite
chirality for each color structure, then a positive $\afb$ can
be induced through either left-handed up quarks and
right-handed top quarks or vice versa. In the vector-axial
basis defined in Eq.~\eqref{AV_ops} this corresponds in
particular to $c_V^8=-c_A^8\,$, affecting at the interference
level both the $t\bar t$ cross section and the asymmetry.
Indeed we find that a similar $\chi^2$ fit in the chiral case
is worse by at least 2$\sigma$ than the best fit point in the
general case. For illustration, at the minimum of $\chi^2$ in
the chiral case, we find $\afb\lesssim 20\%$ and $\sim 2\sigma$
tension with the cross section.

In principle, this analysis does not rely on any specific UV
completion. Yet, specific models of heavy NP can be mapped to
the EFT description of Section~\ref{sec:ops}. To this end, we
use the list of models presented in
Ref.~\cite{delAguila:2010mx}, considering the addition of each
new particle individually. (These models were examined in the
context of the forward-backward asymmetry in
\eg~\cite{AguilarSaavedra:2011vw}.) Since most of the models
contribute to the set of EFT operators via a single chirality
structure, we find that they are in tension with the data, as
mentioned above. The only model of a single new particle that
can give independent values to $c_A^8$ and $c_V^8\,$, therefore
providing a good match to the fit, is of a heavy color octet
and $SU(2)_L$ singlet, namely an axigluon.

\section{LHC Predictions and Constraints} \label{sec:lhc}

We now discuss predictions for LHC measurements arising from
the EFT description of the top Tevatron data. We first compare
the above fit results with existing LHC data based on the 7~TeV
run, and then provide predictions for the 2012 run at 8~TeV.

Regarding the $t \bar t$ cross section at the LHC, the heavy NP
explanation of the Tevatron data can significantly affect the
top pair spectrum only at invariant masses above
$\sim1$~TeV~\cite{Delaunay:2011gv}. In this regime, top decay
products are highly collimated such that they merge into a
single jet, and their reconstruction therefore requires
dedicated jet substructure techniques. Currently, there is no
partonic level result for the shape of the $t\bar t$ spectrum
in this boosted regime. However, CMS has recently published an
upper bound on the enhancement in the $t \bar t$ cross section
integrated over $\mtt>1 \, \mathrm{TeV}$ based on the all
hadronic channel~\cite{Chatrchyan:2012ku}:
\beq \label{Sbound}
\mathcal{S} \equiv \frac{\int_{\mtt>1\,\mathrm{TeV}} \frac{
\mathrm{d} \sigma_{\mathrm{SM+NP}}}{\mathrm{d}\mtt}
\mathrm{d}\mtt}{ \int_{\mtt>1\,\mathrm{TeV}} \frac{ \mathrm{d}
\sigma_{\mathrm{SM}}}{\mathrm{d}\mtt} \mathrm{d}\mtt} <2.6 \,,
\eeq
at 95\% confidence level. This bound can be used to constrain
the EFT parameter space, as depicted in Fig.~\ref{fig:chi2fit}.
It is evident that the bulk of the parameter space that
accounts for the Tevatron data is left intact, with a
prediction of $\mathcal{S}\simeq 1.8$ for the best fit point
reported above. Future improvement of this measurement might
further constrain the EFT parameters.

\begin{figure}[tb]
  \centering
  \includegraphics[scale=0.8]{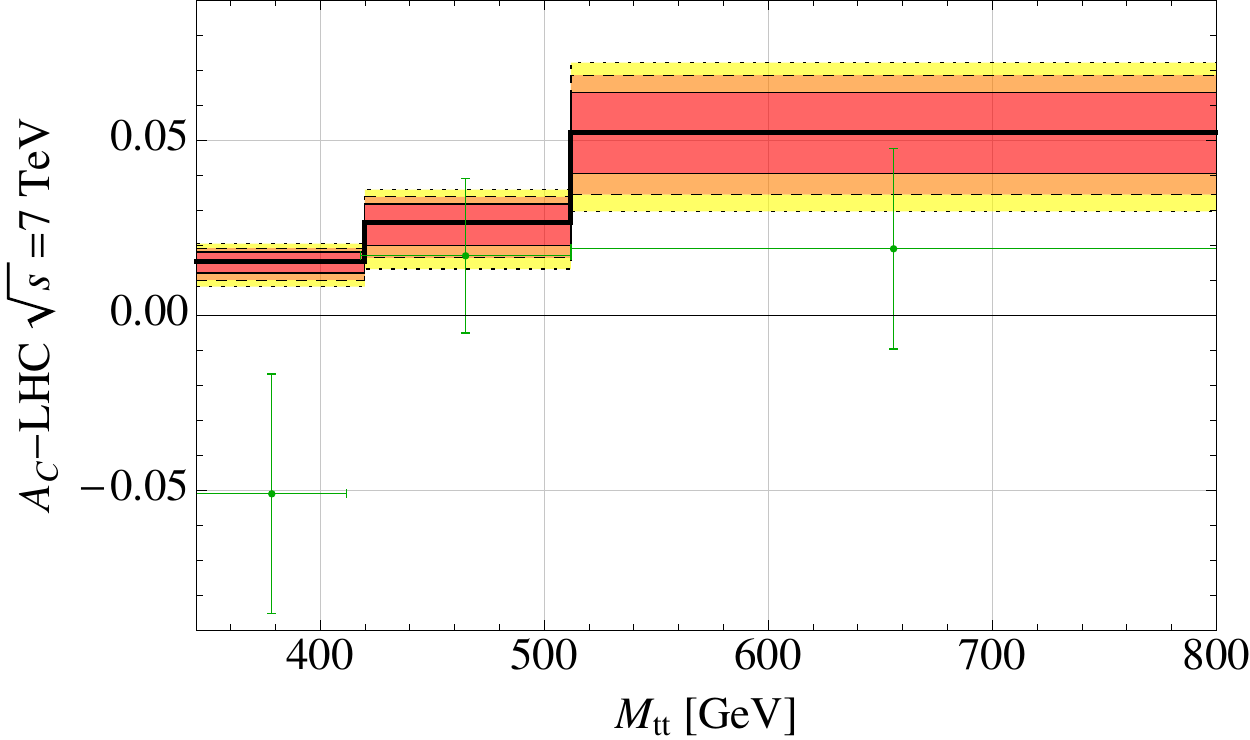} \includegraphics[scale=0.8]{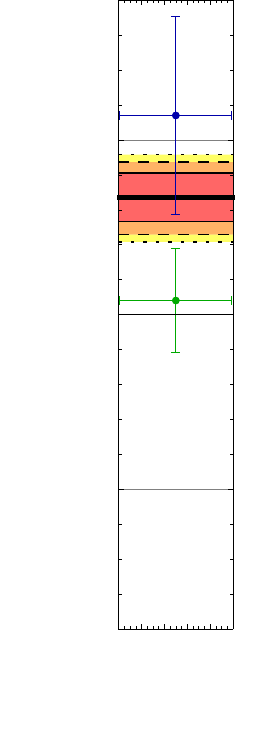}
  \caption{Comparison of the EFT fit to the differential (left) and inclusive (right)
  charge asymmetry to the LHC data. The red, orange and yellow shaded regions correspond
  to 1$\sigma$, 2$\sigma$ and 3$\sigma$ of the global fit to Tevatron data,
  respectively, and the green points and error bars describe the CMS measurement~\cite{cmsAC}.
  The inclusive charge asymmetry measured by ATLAS~\cite{atlasAC} is shown in blue.
  \label{fig:AC7}}
\end{figure}

Next we consider the measured charge asymmetry in
Eq.~\eqref{ACdata}, including the differential data as a
function of $\mtt\,$. Fig.~\ref{fig:AC7} presents the charge
asymmetry stemming from the fit, compared to the
CMS~\cite{cmsAC} and ATLAS~\cite{atlasAC} data.  Currently,
there is no strong tension between the Tevatron measurement of
the top forward-backward asymmetry and the LHC charge asymmetry
data within the EFT framework. It will be interesting to see
how future updates of the data affect the EFT paradigm.

In accordance with the above results, we have verified that
including the charge asymmetry and $t \bar t$ enhancement data
in the fit affects the result only mildly, leaving the
conclusions unchanged.

\begin{figure}[thb]
  \centering
  \includegraphics[width=.6\textwidth]{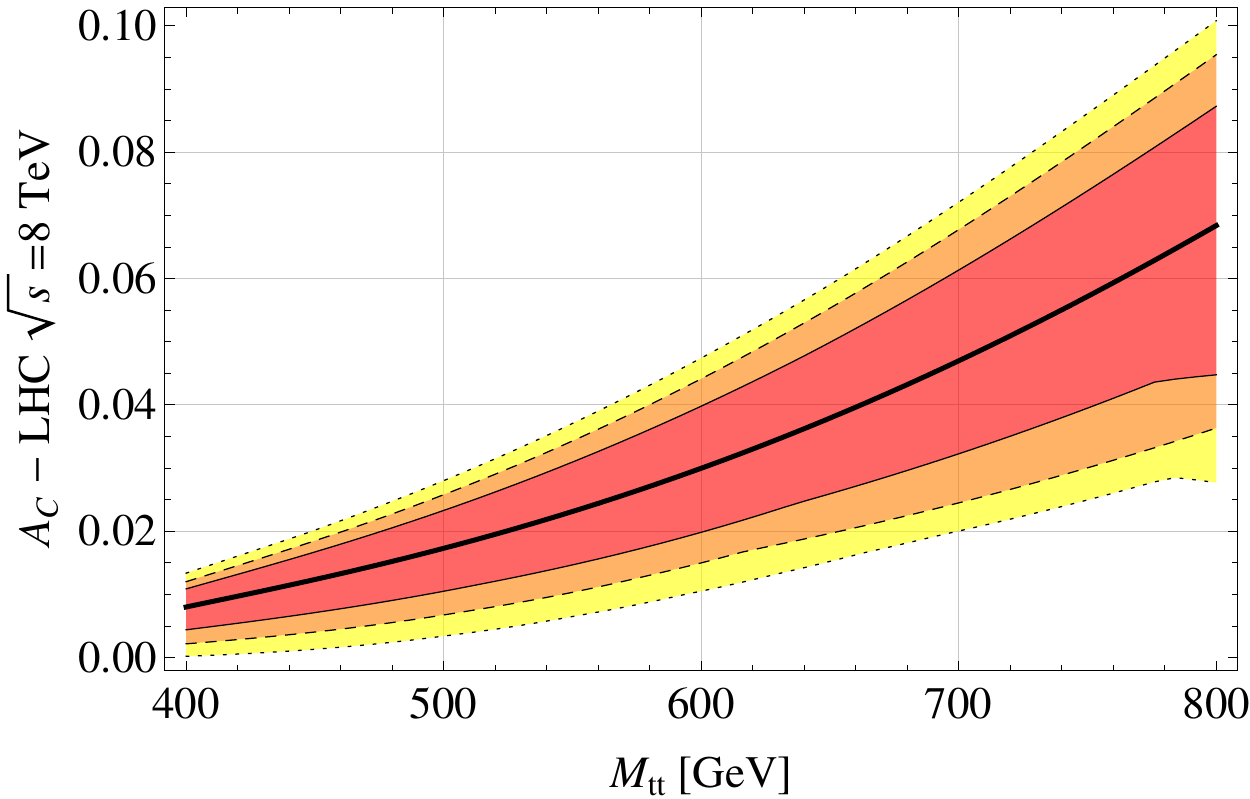}
  \caption{Prediction for the NP contribution (not including the SM) to the differential charge
  asymmetry as a function of $\mtt$ for the 8~TeV LHC run. The red, orange and yellow shaded
  regions correspond to 1$\sigma$, 2$\sigma$ and 3$\sigma$ of the global fit to Tevatron data,
  respectively. \label{fig:AC8}}
\end{figure}

\begin{figure}[thb]
  \centering
  \includegraphics[width=.6\textwidth]{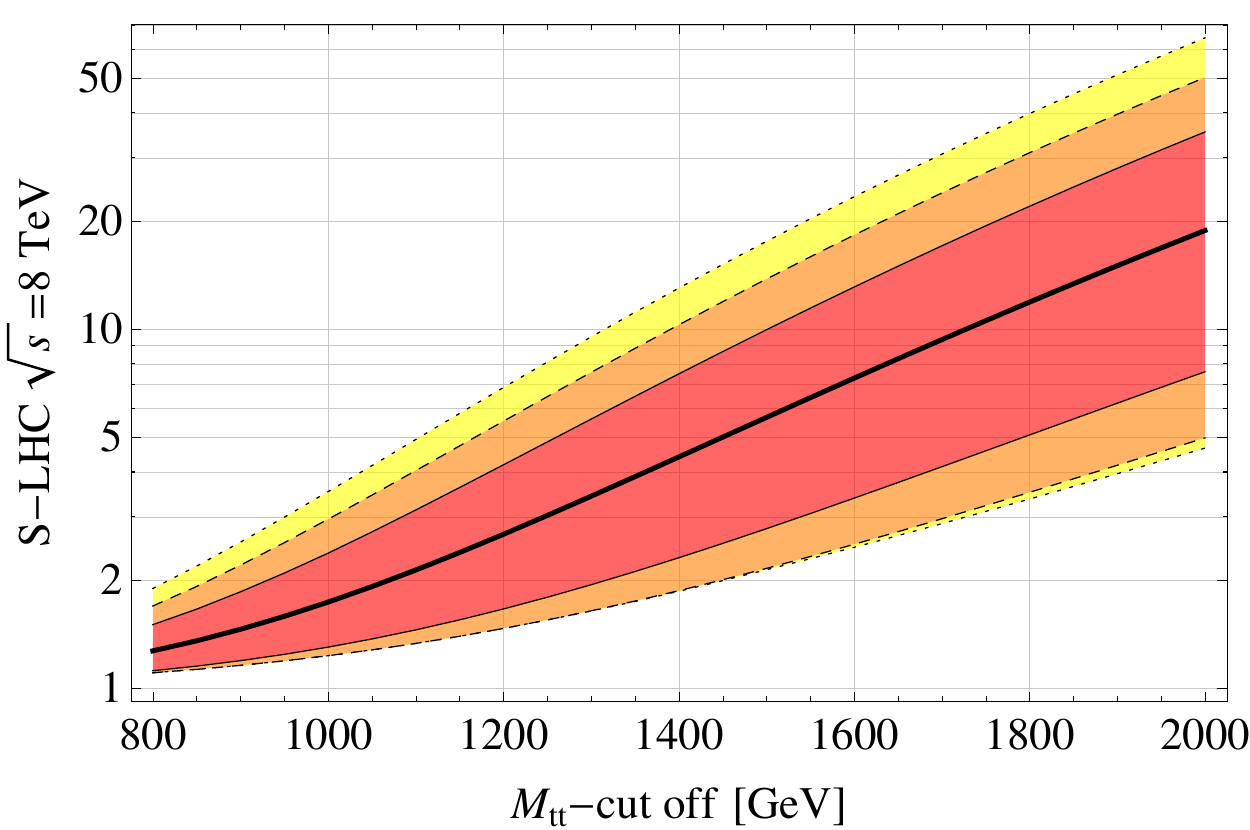}
  \caption{Prediction for the enhancement of the differential $t \bar t$ cross section as a
  function of the cutoff $\mtt$ for the 8~TeV LHC run, parameterized by the appropriate modification of Eq.~\eqref{Sbound}.
  The red, orange and yellow shaded regions correspond to 1$\sigma$, 2$\sigma$ and 3$\sigma$
  of the global fit to Tevatron data, respectively. \label{fig:S8}}
\end{figure}

The EFT parameter space that fits the data can also be used to
give predictions for future LHC measurements at 8~TeV. We focus
on two observables: the differential charge asymmetry and the
differential $t \bar t$ cross section. In Fig.~\ref{fig:AC8} we
show the NP contribution to the former, while Fig.~\ref{fig:S8}
presents the latter, as described by a modified $\mathcal{S}$
parameter as a function of the cutoff $\mtt\,$, along the lines of Eq.~\eqref{Sbound}. We learn that the
heavy NP explanation for the top forward-backward asymmetry
predicts an enhancement of the $t \bar t$ differential cross
section at high $\mtt\,$, as was first pointed out
in~\cite{Delaunay:2011gv}. For example, for $\mtt>1.5$~TeV, the
minimal enhancement is a factor of 2 above the SM cross section prediction.
We also learn that, in the EFT context, the Tevatron
$\afb$ predicts a positive $A_{\rm C}$ which grows with energy
up to $\mtt\simeq800$~GeV.

\section{Dijet Constraints} \label{sec:di}

Dijets searches are sensitive to NP dynamics with sizable
couplings to light flavors or gluons. Since agreement with the
expected SM rate has been observed thus
far~\cite{Chatrchyan:2012bf,atlasdijets}, the available LHC
dijet data provides constraints on NP models explaining the top
$\afb\,$. In the context of heavy NP, the most naive
interpretation of the operator basis defined in
Section~\ref{sec:ops} does not generate a dijet signal from a
single operator, as the top quark in general escapes the cuts
used in the relevant measurement.

\begin{figure}[tb]
  \centering
  \includegraphics[width=.6\textwidth]{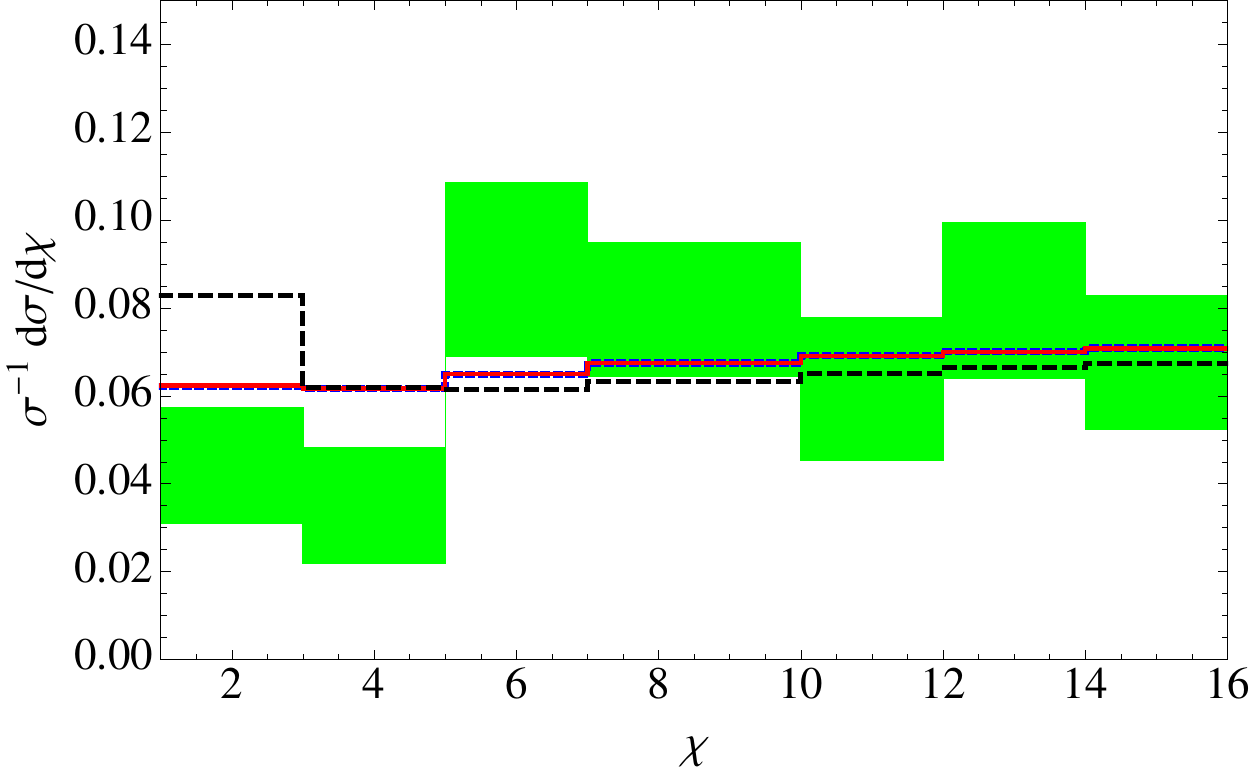}
  \caption{Various contributions at leading order and at the partonic level to the angular distribution of dijet
  production with $M_{jj}>3$~TeV: The SM (dotted blue); heavy NP with
  $c_A^8=1.35\,$~TeV$^{-2}$ and all other operators set to zero (solid red);
  and the limiting operator $+2\pi/(8\,{\rm TeV})^2(\bar q_L \gamma^\mu q_L)^2$
  of~\cite{Chatrchyan:2012bf} (dashed black). The measured 1$\sigma$ bands
  are depicted in shaded green. \label{fig:di}}
\end{figure}

If the underlying theory is $SU(2)_L$-invariant, then in a
chiral basis couplings between first generation quarks and
left-handed $b$ quarks are automatically present. We
conservatively assume that the data does not differentiate
between the various (non top) quark flavors, such that $b$ jets
are bounded by the current dijet data. Our procedure is as
follows. Any point in the fit parameter space can be projected
onto the chiral basis, where left handed tops and $b$'s are
interchangeable (we consider both singlet and triplet EW
contractions). The angular distribution of the resulting $u
\bar u \to b \bar b$ contribution to dijet production can then
be calculated and compared to the data. In Fig.~\ref{fig:di} we
demonstrate this using $c_A^8=1.35$~TeV$^{-2}$ and all other
operators turned off, which is representative of the parameters
best fitting the Tevatron data. We use a singlet EW
contraction, and have checked that a triplet contraction (as
well as adding $d\bar d$ contributions) does not alter the
results. The contribution to dijet production using the CMS
method~\cite{Chatrchyan:2012bf} is shown at the partonic level
along with the measurement and the SM contribution to the high
dijet invariant mass region, $M_{jj}>3$~TeV. Note that this
analysis is only valid for a heavy NP scale above 3~TeV,
otherwise a model-specific calculation is required. Also shown
is the distribution coming from the operator $(\bar q_L
\gamma^\mu q_L)^2$, where $q_L$ denotes the first generation
quark left-handed doublet, taken at its limiting value as found
in~\cite{Chatrchyan:2012bf}. It is evident that the heavy NP
contribution tightly follows that of the SM, and is
significantly below the limiting contribution that defines the
bound. Hence, this analysis leaves the EFT parameter space
unconstrained.

At the loop level, dijets are unavoidably produced by moving to
a chiral basis and exchanging a $W$ between left-handed tops,
resulting in an effective coupling to down type quarks. The
dominant contribution to this dijet signal comes from final
state $b$ quarks. As this signal is necessarily smaller than
that of the unconstrained tree level process discussed above,
we conclude that there are no model-independent dijet
constraints on the heavy NP parameter space.

If the operators emerge from an s-channel exchange of a new
heavy color octet particle with different axial couplings to up
($g_{uu}$) and top quarks ($g_{tt}$), then dijets are
unavoidably produced~\cite{Delaunay:2011gv}. Combining the top
$\afb$ and dijet data sets a bound on the ratio of the
couplings:
\beq\label{dijethierarchy}
g_{uu}/g_{tt}\lesssim 1/6\,.
\eeq
In deriving Eq.~\eqref{dijethierarchy} we use the latest CMS
bound reported for an axial color singlet
operator~\cite{Chatrchyan:2012bf}, where the operator is
rescaled by a factor of $1/N_c$ to account for the different
color contraction.

The simplest version of the above NP model is the case of a
heavy axigluon. Such a colored octet state with a mass $m_A$
and purely axial couplings to quarks induces only the $\co_A^8$
operator at low energies, and can easily accommodate the
preferred data region in Fig.~\ref{fig:chi2fit} with $c_A^8
\equiv g_{uu}\,g_{tt}/m_A^2 \simeq 1 \, \mathrm{TeV}^{-2}$.
This can be achieved with $\co(1)$ perturbative couplings also
satisfying the dijet constraint of Eq.~\eqref{dijethierarchy}
and mass $m_A$ of order a few TeV, above the present reach of
the LHC. We have also verified that a lighter axigluon with
$m_A \sim2$~TeV, which can be produced on-shell at the LHC, is
not excluded by the dijet data (although the ratio of its
couplings $g_{uu}/g_{tt}$ must be somewhat smaller than the
upper bound in Eq.~\eqref{dijethierarchy} in order to account
for the Tevatron $\afb$).

Dijets can also arise by sewing together two operators to
generate a $u \bar u \to u \bar u$ process through a top
loop~\cite{Domenech:2012ai}. However, this loop is
quadratically divergent, and so the corresponding constraints
are highly model-dependent. For instance, if the
$\mathcal{O}_A^8$ operator arises from the exchange of a
massive spin one state, gauge invariance would protect against
the quadratic divergence.

Similarly, contributions to atomic parity violation and $Z \to
b \bar b$ can emerge through quadratically divergent top and up
loops, respectively. In addition to these processes being
model-dependent, they can only stem from color-singlet
operators, which have a mild impact on the above results.

\section{Implications for $b \bar b$ Production Asymmetry}
\label{sec:bb}

We now discuss the effect of the EFT operators considered in
this work on the forward-backward asymmetry in $b \bar b$
production. If the underlying theory is $SU(2)_L$-invariant
then couplings between first generation quarks and left-handed
$b$ quarks necessarily emerge, as mentioned in the previous
section.

\begin{figure}[tb]
  \centering
  \includegraphics[width=.6\textwidth]{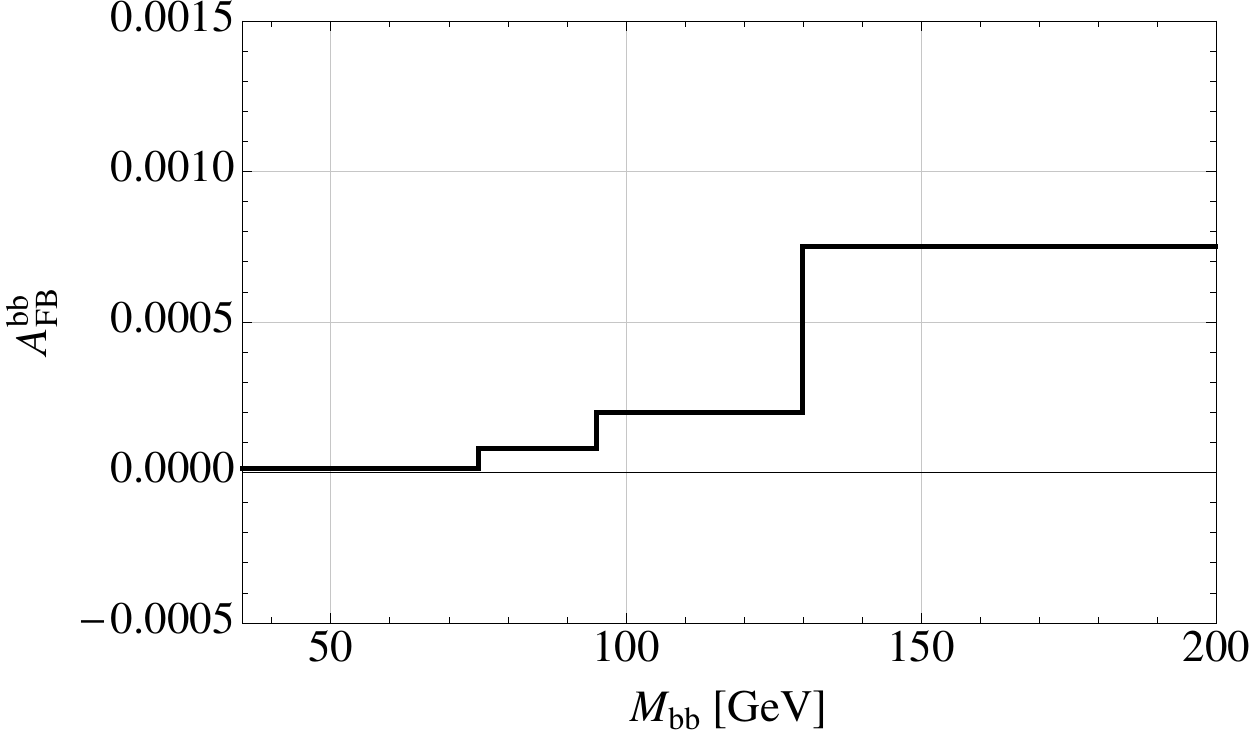}
  \caption{Prediction for the NP contribution to $b \bar b$ forward-backward asymmetry for
  $c_A^8=1.35\,$~TeV$^{-2}$ and all other operators set to zero. \label{fig:bb}}
\end{figure}

In Fig.~\ref{fig:bb} we show the maximal forward-backward
production asymmetry of the $b\bar b$ system, $A_{\rm
FB}^{b\bar b}\,$, that can be obtained for
$c_A^8=1.35\,$~TeV$^{-2}$ and all other operators set to zero.
This is obtained with an $SU(2)_L$ singlet contraction; the
triplet contraction leads to a smaller asymmetry. For other
points in the fit region the relative change in the result is
at most of $\co(1)$. The asymmetry is plotted as a function of
$M_{b \bar b}$ according to the binning reported by
CDF~\cite{CDFbb}.

We find that the NP contribution to the asymmetry at invariant
mass of up to 200~GeV cannot exceed a permil. In particular, a
measurement of $A_{\rm FB}^{b\bar b}$ of order a percent would
be inconsistent with this prediction. We conclude that if a
large $A_{\rm FB}^{b\bar b}$ is measured, a heavy NP theory
which accounts for the top forward-backward asymmetry must
include an additional source for the bottom asymmetry beyond
the $SU(2)_L$-completion of the operators of
Eqs.~\eqref{AV_ops} and~\eqref{ST_ops}.

\section{Conclusions} \label{sec:conc}

We have performed a model independent analysis for the $t \bar
t$ forward-backward asymmetry and other top-related
measurements, assuming heavy new physics. We have shown that
this framework is still largely consistent with all the
available data. We provide robust predictions for both LHC and
Tevatron observables, which will be tested in the near future.
Specifically, we predict an enhancement at high invariant
masses of both the top pair production cross section and the
charge asymmetry at the LHC. Additionally, the $b \bar b$
forward-backward asymmetry at the Tevatron cannot exceed
$\co(10^{-3})$ at invariant mass of up to 200~GeV.

The 2012 LHC run should be able to probe the parameter space of
the heavy new physics. In particular, since the existing CMS
measurement of the differential cross section already places
partial constraints on the heavy new physics hypothesis, we
expect the 8~TeV data to improve the bounds. This will require
the use of dedicated jet reconstruction techniques, since for
high invariant masses the top quark is highly boosted, and its
decay products typically merge into a single (fat) jet. These
new techniques are showing constant improvement (see
\eg~\cite{Chatrchyan:2012ku,Aaltonen:2011pg,ATLAS:2012am,Aad:2012np}),
and will soon be able to provide a partonic level measurement
of the $\mtt$ distribution in the boosted regime, where TeV
scale new physics might be at play.

\section*{Acknowledgments}
We thank Christophe Grojean, Alex Kagan, Gilad Perez and Ofer
Vitells for useful discussions, and Yossi Nir
for comments on the manuscript.

\end{document}